\documentstyle[12pt]{article}
\begin{document}

\newcommand\ie {{\it i.e. }}
\newcommand\eg {{\it e.g. }}
\newcommand\etc{{\it etc. }}
\newcommand\cf {{\it cf.  }}
\newcommand\etal {{\it et al. }}
\newcommand{\be}{\begin{eqnarray}}
\newcommand{\ee}{\end{eqnarray}}
\newcommand\Jpsi{{J/\psi}}
\newcommand\M{M_{Q \overline Q}}
\newcommand\mpmm{{\mu^+ \mu^-}}
\newcommand{\jp}{$ J/ \psi $}
\newcommand{\pp}{$ \psi^{ \prime} $}
\newcommand{\ppp}{$ \psi^{ \prime \prime } $}
\newcommand{\dd}[2]{$ #1 \overline #2 $}
\newcommand\noi {\noindent}
\bibliographystyle{try2}

\begin{flushright}
LBL-35380\\
SLAC-PUB-6468 \\
April 1994 \\
(T/E)
\end{flushright}
\vspace{1cm}

\begin{center}

{\Large { QCD and Intrinsic Heavy Quark Predictions for Leading Charm
and Beauty Hadroproduction$^{\star}$}
}\\[7ex]

R. Vogt
\\[2ex]

Lawrence Berkeley Laboratory\\
Berkeley, California\quad 94720 \\[2ex]

and \\ [2ex]

S. J. Brodsky
\footnotetext{$\star$ Supported in part by the
U. S. Department
of Energy under Contract Numbers DE-AC03-76SF00515 and
DE-AC03-76SF0098.} \\[2ex]

Stanford Linear Accelerator Center\\
Stanford University\\
Stanford, California\quad 94309 \\[3ex]

\vspace{1cm}
\begin{quote} \begin{small}
Recent experiments at Fermilab and CERN
have observed a strong asymmetry  between the
hadroproduction cross sections of leading $D$
mesons, containing projectile valence quarks,
and  nonleading charmed mesons,
without projectile valence
quarks.  The observed correlations of the $\pi^{\pm} N \to D^\pm X$
cross section
with the projectile charge  violates the usual assumption that
heavy quark jet
fragmentation factorizes.
We examine the asymmetry between leading and nonleading charm
production as a function of $x_f$ and $p_T^2$ assuming
a two-component model
combining leading-twist fusion subprocesses and
charm production from intrinsic
heavy quark Fock states.  We predict a sizable asymmetry at low
$p_T^2$ and high $x_f$ from coalescence of the charm quarks
with the comoving
spectator quarks of the projectile.
An intrinsic $c \overline c$ production
cross section of 0.5 $\mu$b is sufficient to
explain both the magnitude and
kinematic dependence of the asymmetry.  In contrast, the charm jet
hadronization
mechanisms contained in PYTHIA
predict a sizeable leading charm asymmetry even at low $x_F.$
The two-component model is extended to predict the asymmetry in
$B$ meson production in proton-proton and pion-proton interactions.
\end{small}
\end{quote}
\end{center}
\newpage
\begin{center}
{\bf I. Introduction}\\[2ex]
\end{center}

In leading-twist QCD, the factorization theorem \cite{fact}
predicts that the
fragmentation functions $D_{H/c}(z,Q)$ are independent
of the quantum numbers
of both the projectile and target.
However,   strong flavor correlations between the
produced particle and the projectile have been reported
in charm production
\cite{Agb,Bar}.
For example, in $\pi^- (\overline u d)$ interactions with
hadrons or nuclei, the $D^- (\overline c d)$ $x_f$
distribution is consistently
harder than
the $D^+ (c \overline d)$ distribution.
The $D^-$ and $D^0 (c \overline u)$
are referred to as ``leading"  charmed mesons
while the $D^+$ and $\overline
D^0 (\overline c u)$ are ``nonleading".
This leading behavior thus suggests
that hadronization at large $x_F$ involves the
coalescence of the produced
charm or anticharm quarks with the spectator quarks of the projectile,
just
as in exclusive reactions.
The study of this phenomena thus can provide new
insights into the coherent mechanisms
controlling the formation of hadrons in QCD.

Charmed hadron distributions are often parameterized
as $\propto (1-x_f)^n$, where $n_{\rm nonleading}$
is observed to be larger
than $ n_{\rm leading}$ \cite{Agb,Bar}.
However, this kind of parameterization
is rather insensitive to the details of
the  mechanisms responsible for leading
charm effects.
A more sensitive observable, the asymmetry between leading and nonleading
charm, has been used in the recent
analyses of the WA82 \cite{WA82} and E769
\cite{E7692} collaborations.  The asymmetry, defined as
\be {\cal A} =
\frac{\sigma(\rm leading) - \sigma(\rm nonleading)}{\sigma(\rm leading) +
\sigma(\rm nonleading)} \, \, , \ee
does not require the assumption of the functional
form of the cross section.
Both experiments find that the measured asymmetry
${\cal A}(x_f),$ integrated over $p_T$,
increases from $\sim 0$ for $x_f$ near
zero to $\sim 0.5$ around $x_f = 0.65$
\cite{WA82,E7692}.
However, the asymmetry ${\cal A}(p_T^2),$  integrated over
all $x_f$, is found to be consistent
with zero in the range
$0<p_T^2< 10$ GeV$^2$ \cite{E7692}.  These facts are consistent if
the leading charm asymmetry is localized at large $x_F,$ involving
only a small fraction of the total cross section.

The experiment
WA82 measures charmed hadron
production by 340 GeV $\pi^-$ beams on W/Si and W/Cu targets \cite{Adam}.
Experiment E769 has measured $D^\pm$
production in Be, Al, Cu, and W targets
\cite{E769} with a mixed beam (75\%
$\pi^-$ and 25\% $\pi^+$) at 250 GeV.  The
$D^+$ is leading for a $\pi^+$ projectile.
In each experiment only $D^\pm$
mesons are used in the asymmetry analysis to avoid ambiguities in the
assignment of leading and nonleading charm to $D^0$ mesons.  The $D^0$, a
leading charm state when directly produced by $\pi^-$ beams, can also be
produced by $D^{+ \star}$ and $D^{0 \star}$ decays; {\it e.g.} 50\% of
nonleading $D^{+ \star}$'s decay to $D^0$.

Perturbative QCD at leading order predicts that
$c$ and $\overline c$ quarks are produced
with identical distributions.  Next-to-leading order calculations do give
rise to a small charge asymmetry ($\sim$ 10\% for $x_f \sim 0.8$) between
$\overline c$ and $c$ production due to $qg$ and $q \overline q$
interference \cite{Been,NDE}.  However, this charge asymmetry should
result in an increase of $D^-$, $\overline D^0$ production over $D^+$ and
$D^0$ at high $x_f$, not a separation between $D^-$, $D^0$ and $D^+$,
$\overline D^0$.

How can one explain the origin of
leading charm asymmetry within the context of
QCD? It
is clear that the produced charm (or anticharm) quark must combine with a
projectile valence
quark.  Ordinary jet fragmentation ({\it e.g.}\ Peterson fragmentation
\cite{Pete}) cannot
produce a leading particle asymmetry
since it is independent of the initial
state and thus the projectile valence quarks.
This is an essential property of
leading-twist factorization. However, one expects on physical
grounds that a charm quark produced by fusion
may coalesce with a comoving
spectator
valence quark \cite{hwa,bg,bedny,andr}.  For example, in QED, leptons of
opposite charge moving with similar velocities
can be captured into neutral
atoms \cite{bgs}.  Since the capture is significant only at
small relative rapidity, $\Delta y$, the effect on the total rate is
higher twist.

In leading-twist QCD heavy quarks are produced by the
fusion subprocesses $g g \rightarrow Q \overline Q$ and $q \overline q
\rightarrow Q \overline Q$.   The heavy
$Q$ or $\overline Q$  normally fragments independently;  however,
there is a finite probability that
it will combine with a spectator valence quark in the final state
to produce a leading hadron. Coalescence is expected to
dominate when the valence quark
and the produced heavy quark have the same
velocity. The coalescence amplitude should be largest at small relative
rapidity
since the invariant mass of the $\overline Q q $ system is minimal
and the binding
amplitude of the heavy meson wavefunction is maximal. This picture of
coalescence is also consistent with ``heavy quark
symmetry" \cite{isw,cale}.
A similar final-state coalescence mechanism is contained in PYTHIA,
a Monte Carlo program based on the Lund string
fragmentation model \cite{PYT}.  Its string mechanism produces some
charmed hadrons with a substantially larger
longitudinal momentum than the charmed quarks originally
produced by the fusion processes.
At large $x_f$ and low invariant string
masses, the produced $D^-$ or $D^0$ inherits all the remaining projectile
momentum while $D^+$, $\overline D^0$ production is forbidden.  However,
PYTHIA substantially overestimates the observed asymmetry
${\cal A}(x_f),$  particularly
at low $x_f$.  It also results in ${\cal A}(p_T^2) \sim 0.3$ for
$0<p_T^2<10$ GeV$^2$, overestimating the effect seen in the
$x_f$-integrated data.

The difficulty that PYTHIA has in
reproducing the shape of the leading charm
asymmetry seems to be a characteristic  feature of models based on
final-state coalescence.  If we parameterize the fusion-produced charm
distribution
as $dN_c/dx \propto (1-x)^n,$ then the leading charmed hadron
spectrum falls more slowly due to the extra momentum supplied by
the
projectile quark $dN_D/dx_f \propto
[1- x_f (1-\delta)]^n$ where $\delta \sim
m_{T_{\overline q}}/m_{T_D}$ and $m_T$ is the transverse mass.
However, the asymmetries generated by this
parameterization rise faster than the observation at low $x_f$ that
${\cal A}(x_f)$ is negligible  when $x_f  < 0.25$ \cite{WA82,E7692}.
Thus a coalescence model strong enough
to reproduce
the hard spectrum of the leading charmed
hadrons starting from the relatively
soft fusion-produced charmed quarks tends to produce an
excessivly large leading charm asymmetry over all phase
space.
It would be illuminating if the (true) rapidity
distributions of the leading and nonleading charmed hadrons were measured
to see whether or not their
rapidity distributions track those of the underlying
fusion-produced charmed quarks.

In the above picture of leading charm hadroproduction,
it is implicitly assumed
that coalescence is strictly a final-state phenomenon. In fact,
the coalescence of the charm quark and a projectile valence quark may
also occur in the initial state.  For example, the
$\pi^-$ can fluctuate into a
$| \overline u d c \overline c \rangle$ Fock state.  The most important
fluctuations occur at minimum invariant mass ${\cal M}$ where all the
partons have approximately the same velocity.   Characteristically,
most of the momentum is carried by the heavy quark
constituents of these Fock states.
As viewed from the target rest frame, the intrinsic charm
configurations can have very long  lifetimes, of order $\tau = 2 P_{lab}/
{\cal M}^2$ where $P_{lab}$ is the projectile momentum.
Intrinsic charm hadroproduction occurs dominantly when the spectator
quarks interact strongly in the target \cite{BHMT}, explaining
why large $x_f$ charm production on nuclear targets is observed to have
a strong nuclear dependence, similar to that of the inelastic
hadron-nucleus cross section.
Since the charm and valence quarks have the
same rapidity in an intrinsic charm Fock state, it is easy for
them to coalesce into charmed hadrons and produce leading particle
correlations at large $x_f$ where this mechanism can dominate the
production rate. This is the basic
underlying assumption of the intrinsic charm
model \cite{intc}.

The leading charm asymmetry must be a higher-twist effect or
it would violate pQCD
factorization.  Final-state coalescence is higher twist since only a
small fraction of the fusion-produced heavy quarks will combine with
the valence
quarks.  Intrinsic heavy quark production is
also higher twist--because the virtual configurations in the
projectile wavefunction must be
resolved during their limited lifetime.  The
cross section decreases with extra powers of $1/m_Q$ relative to
leading-twist fusion.  From a general quantum-mechanical standpoint, both
types of higher-twist mechanisms,
coalescence of fusion-produced charm in the
final state and coalesence of the intrinsic charm
configurations in the initial state, must occur in QCD at some level.

In this paper, we shall calculate
the asymmetry within a two-component model:
parton fusion with coalescence, and intrinsic charm with valence-quark
recombination \cite{VBH2}.\\[3ex]

\begin{center}
{\bf II. Leading-Twist Production}\\[2ex]
\end{center}

The inclusive cross section for a single charmed hadron as a function of
$x_f$, $x_f = (2m_T/\sqrt{s}) \sinh y$, and $p_T^2$ in leading twist
QCD has the factorized form \cite{VBH2}
\be
\sigma = \frac{\sqrt{s}}{2} \int H_{ab}(x_a,x_b) \frac{1}{E_1}\
\frac{D_{H/c}(z_3)}{z_3}\ dz_3 dy_2 dp_T^2 dx_f \, \, ,
\ee
where $E_1$ is the energy of the
charmed quark, $y_2$ is the rapidity of the
charmed antiquark, $H_{ab}(x_a,x_b)$
is the convolution of the differential
cross section with the parton distribution
functions, and $m_c = 1.5$ GeV. We
shall use the lowest-order parton fusion calculation.
A $K$ factor of $\sim
2-3$ is included in the normalization of the fusion cross section. For
consistency with the LO approximation,
we use two current leading order sets of
parton distribution functions, GRV LO \cite{GRV} and Duke-Owens 1.1
\cite{Owens} for the proton, and their
pion counterparts, GRV LO \cite{GRVpi}
and Owens set 1 \cite{Owenspi}.

The fragmentation function, $D_{H/c}(z_3)$,
describes the hadronization of the
charmed quark where $z_3 = x_D/x_c$ is the fraction of the charm momentum
carried by the charmed hadron, assuming it
is collinear with the charmed quark.
 We have studied two different fragmentation functions, a delta function,
$\delta(z_3 -1)$,
and the Peterson function extracted from $e^+e^-$ data \cite{Pete}.  We
have shown previously that the Peterson function
predicts a softer $x_f$
distribution than observed in hadroproduction,
even at moderate $x_f$.  The
delta function model assumes that the  charmed quark coalesces with
a low-$x$ quark spectator from the sea
(or a low momentum secondary quark produced in the collision)
so that the charmed quark retains its
momentum and velocity \cite{VBH2}.  Either choice of
fragmentation function  is
independent of the initial state and does not produce
flavor correlations between the projectile
valence quarks and the final-state
hadrons.

In Fig.\ 1 we show the $x_f$ distributions
calculated for (a) $\pi^- p$ and (b)
$\pi^+ p$ interactions at 250 GeV and (c)
$\pi^- p$ interactions at 340 GeV
using each set of structure functions with both choices of charm quark
fragmentation function.  The solid and dot-dashed
curves give the calculated
distributions using delta function and Peterson
function fragmentation with GRV LO.
The dashed and dotted curves illustrate
the same calculations with DO 1.1.
For charm production at these energies,
DO 1.1 gives a somewhat larger cross
section. As expected, delta function fragmentation results in harder
distributions than those predicted by
Peterson fragmentation for $x_f > 0.2$.
Note however that the conventional fusion model, even with delta function
fragmentation cannot account for the
shape of the leading $D$ production cross
section from WA82 \cite{WA82}.
(Since the normalization of the data has not
been fixed, we normalize it to our
calculated cross section.)  The differences
in the fusion cross sections from the GRV and DO distributions are only
apparent at low $x_f$ where $gg
\rightarrow c \overline c$ dominates since the
gluon distributions are uncertain.  At large $x_f$, where $q \overline q
\rightarrow c \overline c$ is more important,
little difference can be observed
in the $x_f$ distributions because the valence
distributions are relatively
well measured.
In the following we will use the GRV LO distributions only.

For valence quark coalescence
to be effective as a leading charm production
mechanism, either the
$c$ or $ \overline c$  must be comoving with the projectile
valence quarks since capture into a bound state wavefunction favors
constituents with similar rapidities.
Is this possible in the fusion model? Most of the produced
charmed quarks produced in the fusion reaction
presumbably hadronize into charmed mesons or baryons
independent of the projectile identity.
To first approximation, the spectator quarks have the
same rapidity as the projectile itself since they
are bound state components of the projectile wavefunction.
The produced heavy quarks tend to have low rapidity compared to the pion
valence quarks.
In the Lund string fragmentation model,
as contained in PYTHIA \cite{PYT},
a charmed quark is always found at the endpoint of a string.  This string
pulls the charmed quark in the direction
of the other string endpoint which
is usually a beam remnant.  When the two
string endpoints are moving in the
same general direction, the charmed hadron can then be produced at larger
longitudinal momentum that then original charmed quark.  In the extreme
case where the string invariant mass is too small to allow the production
of several particles, the string scenario reduces to a coalescence one,
with the two string endpoints determining the flavor content of the
produced hadron \cite{torb}.  In Fig.\
2(a) we compare the pion valence and sea
quark rapidity distributions with the
fusion-produced charmed quark rapidity distribution at 340 GeV.
The solid and dotted curves are the pion valence and sea quark
distributions while the dashed curve is
the calculated charmed quark rapidity distribution.
At this energy, the kinematic limit for charm production is reached
at $y=2.8.$ The spectator valence quarks tend to have
larger rapidity than the charmed quarks.  In the region where
the distributions overlap, the sea quark rapidity density is also
important.  Thus we shall assume in this paper that
charmed hadrons created from the leading twist fusion
subprocesses arise dominantly either from independent fragmentation,
coalescence with the projectile sea components,  or coalescence
with comoving secondary partons produced in the collision. Therefore
we will also assume that the fusion
mechanism produces a negligible asymmetry between $D^+$ and
$D^-$. We model the coalescence process with delta function
fragmentation.\\[3ex]

\begin{center}
{\bf III.  Intrinsic Heavy Quark Production}\\[2ex]
\end{center}

The fluctuation of a $\pi^-$ into a $|
\overline u d c \overline c \rangle$
Fock state produces a leading particle
asymmetry through recombination of the
intrinsic $c \overline c$ pair with the
comoving valence quarks. The charmed
quarks in the Fock state may be freed
through soft interactions of the light
valence quarks with the target \cite{BHMT}.  The probability distribution
corresponding to an $n$--particle Fock state
(integrated over $k_{\perp i}$) is
assumed to have the form
\be \frac{dP_{\rm ic}}{dx_1 \dots dx_n} = N_n
\alpha_s^4(m_{c \overline c})
\frac{\delta(1-\sum_{i=1}^n x_i)}{(m_h^2 - \sum_{i=1}^n
(\widehat{m}_i^2/x_i) )^2} \, \, ,
\ee
where $\widehat{m}_i = \sqrt{\langle
\vec{k}_{\perp i}^2 \rangle + m_i^2}$ is
the average transverse mass, $\langle
k_\perp^2 \rangle$ is proportional to the
square of the quark mass, and $N_n$ is
the normalization. Here we only consider
the four-particle Fock state\footnote{Introducing
additional light quarks or
gluons reduces both the probability that the pion
will fluctuate into this
configuration and the probability that the
$c \overline c$ will recombine with
a valence quark to produce leading charm due
to the presence of sea quarks.}.
We have assumed the effective values
$\widehat{m}_q = 0.45$ GeV for the valence
quarks and $\widehat{m}_c = 1.8$ GeV for the charm quarks.

The $x$ distribution of intrinsic $c$ quarks in a pion is \be
\frac{dP_{\rm ic}}{dx_c} = \int dx_1 dx_2 dx_{\overline c} \frac{dP_{\rm
ic}}{dx_1 \dots dx_c} \, \, , \ee where $x_1$ and $x_2$ represent the
valence quark momentum fractions.  The intrinsic charm quarks hadronize
both through fragmentation as in the parton fusion model and through
coalesence with valence quarks.  Fragmentation leads to distributions of
the form \be \frac{dP^F_{\rm ic}}{dx_D} = \int dx_c dz_3 D_{D/c}(z_3)
\frac{dP_{\rm ic}}{dx_c} \delta(x_D - z_3 x_c) \, \, , \ee where $x_c$ is
the fraction of the projectile momentum carried by the $c$ quark and the
definition of $z_3$ is identical to that in Eq.\ (2).  The fragmentation
of an intrinsic charm state itself does not produce a leading particle
effect, since $D^+$ and $D^-$ production are equally likely.

The coalescence of one or both of the intrinsic charm quarks with
comoving spectator valence quarks from an intrinsic charm state naturally
produces leading charmed hadrons at large $x_f$.
(By duality, one can think of the leading charmed mesons
as preexisting in the intrinsic heavy quark Fock state.)
The $| \pi^- \rangle =
| \overline u d c \overline c \rangle$ state
may coalesce into $| \overline
u c \rangle$ and $| d \overline c \rangle$,
$D^0$ and $D^-$, automatically
producing leading charm since the charge conjugates contain no valence
quarks.  The leading $D^-$ distribution is calculated from \be \frac{d
P^C_{\rm ic}}{dx_{D^-}} = \int dx_c
 \frac{dP_{\rm ic}}{dx_c} \delta(x_{D^-} - x_d - x_{\overline c}) \, \, ,
\ee where the delta function defines $x_{D^-}$ as the sum of the
$\overline c$ and valence $d$ momentum fractions.  In Fig.\ 2(b) we show
the rapidity distributions for a light quark in a four particle intrinsic
charm state (solid curve) and a charm quark (dashed curve).  The $D^-$
distribution from Eq.\ (6) is also shown (dot-dashed
curve).  The velocity (rapidity)
of a heavy
quark should remain unchanged by hadronization, up to order
$\Lambda_{\rm QCD}/m_Q$.  Thus there should be no acceleration of
the heavy quark.
However, since momentum is conserved, the hadron produced by the
coalescence of equal rapidity partons will have the combined momentum
of the heavy quark and valence spectator
quark: $x_{D^-}=x_q+x_{\overline c}.$  If coalescence dominantly occurs
when the combining partons have the lowest invariant mass,
as in this model, then
$x_q/x_{\overline c} \simeq  m_{T_q}/m_{T_{\overline c}}$.
In contrast, charmed hadrons produced by jet fragmentation have less
momentum than the parent charmed quark.
Thus it is natural that  leading charmed hadrons created by the
coalescence of a
produced quark with a valence quark will have
a harder longitudinal momentum
distribution than nonleading charmed hadrons.
This effect can perhaps also
explain the pattern of leading strange-charmed baryons observed
at large $x_f$ in the CERN WA62 hyperon beam experiment \cite{WA62}.

As in Ref.\ \cite{VBH2}, we assume that the
intrinsic charm model produces
nonleading charm states by fragmentation of
the intrinsic charm states and
leading charm by both fragmentation and valence quark coalescence.
Valence
quark coalescence gives rise to the difference between the leading and
nonleading distributions.  The total charm distribution
is then the sum of the
parton fusion and intrinsic charm components, {\it i.e.} $d\sigma/dx_f =
d\sigma_{\rm pf}/dx_f + d\sigma_{\rm ic}/dx_f$.

Previously we obtained the normalization of the
intrinsic-charm component by
assuming that the ratio of the intrinsic charm
cross section to the total charm
cross section is identical to the ratio of the
``diffractive" to the total
$J/\psi$ production cross section measured by NA3, giving $\sigma_{\rm
ic}/\sigma_{c \overline c}^{\rm total} = 0.18$
for pion-induced production
\cite{Badier}.
The constant $N_4$ was then fixed from this ratio under the
assumption that the probability distribution
and the differential cross section
are identical \cite{VBH2}.
However, because of the uncertainties in relating
$\bar c c$ distributions to quarkonium production,
it is difficult to use the
$J/\psi$ cross section to obtain the absolute
normalization of the intrinsic
charm probability.

We set $N_4$ from the probability $P_{\rm ic}$, with $P_{\rm ic} =
0.31$\% based upon calculations by Hoffmann and Moore \cite{HM} compared
to EMC muoproduction data \cite{EMC} and assume the same $P_{\rm ic}$ for
pion and proton projectiles.  The intrinsic charm cross section will be
proportional to the total inelastic cross section evaluated at $\hat{s}
= (1-x_f)s$ since a soft interaction in the target
breaks the coherence of the Fock state and brings the particles on shell
\cite{BHMT,Chev}.  The cross section thus includes a resolving factor
$\mu^2/4\widehat{m}_c^2$, so that \be \sigma_{\rm ic} = P_{\rm ic}
\sigma_{\pi p}^{\rm in} \frac{\mu^2}{4 \widehat{m}_c^2} \, \, , \ee where
$\mu^2$ is a hadron scale parameter.  Fixing $\mu^2 \sim 0.2$ GeV$^2$
from the NA3 ratio of 0.18 at 200 GeV \cite{Badier}, we obtain
$\sigma_{\rm ic} \sim 1$ $\mu$b for a single charmed hadron with $x_f >
0$ \footnote{Our value of $\sigma_{\rm ic}$ does not contradict the
results of the E653 collaboration from 800 GeV $p$Si interactions
\cite{CMU}, $\sigma_{\rm diff} (D^+)/\sigma(D^+) < 1.8$\%.  At 800 GeV,
using $\sigma_{pp}^{\rm in}$ in Eq.\ (7) and including an $A^{0.71}$
dependence for protons \cite{Badier}, we find $\sigma_{\rm
ic}(D)/\sigma(D) \sim 1.1$\%.}. (The intrinsic $c \overline c$ pair
production cross section is $0.5$ $\mu$b.) The nonleading and leading
distributions are then \be \frac{d\sigma^{NL}_{\rm ic}}{dx_f} & = &
\sigma_{\pi p}^{\rm in} \frac{\mu^2}{4 \widehat{m}_c^2} \frac{dP^F_{\rm
ic}}{dx_f} \\ \frac{d\sigma^{L}_{\rm ic}}{dx_f} & = & \sigma_{\pi p}^{\rm
in} \frac{\mu^2}{4 \widehat{m}_c^2}
\left( (1-\xi) \frac{dP^F_{\rm ic}}{dx_f} +
\xi \frac{dP^C_{\rm ic}}{dx_f} \right)
\, \, , \ee where $\xi$ is a parameter
determining the relative importance of fragmentation and valence quark
coalescence\footnote{We have not made
any distinction between the relative
rates of charged and neutral $D$ production
but have rather assumed the same
proportions for intrinsic charm and fusion.
One could introduce an additional
parameter in an attempt to model production
differences {\it i.e.}\ between
leading $D^-$ and nonleading $\overline D^0$,
both containing $\overline c$
quarks.  Some differences in charged and neutral $D$ production naturally
occur since $D^\star$'s are commonly produced first but do not decay
uniformly since $D^{\star +}$ and $D^{\star 0}$ both decay to $D^0$.  See
Ref.\ \cite{bedny} for some discussion on
relative $D^-$, $D^0$ production by
fusion due to target effects.}.  Following Ref.\ \cite{VBH2},
we have chosen $\xi = 1/2$ in this analysis,
except where otherwise noted.

The nuclear dependence, $A$, of the fusion
model is $A^\alpha$ where $\alpha
\sim 1$ \cite{E7693}. Since intrinsic charm
production tends to occur on the
nuclear surface \cite{BHMT}, its target
dependence is assumed to be $A^\beta$
where $\beta = 0.77$ for pion-induced
reactions \cite{Badier}. The model thus
predicts that leading charm effects are
diminished in nuclear targets because
of the relative suppression of the intrinsic
charm component by $A^{\beta-1}$.
This prediction is consistent with the $A$ dependence seen by the NA3
\cite{Badier} and E789 \cite{E789} collaborations.
We average our calculated asymmetry over
the nuclear targets to compare with
the data.\\[3ex]

\begin{center}
{\bf IV.  Predictions of the Two-Component Model}\\[2ex]
\end{center}

The nonleading and leading $x_f$ distributions
from WA82 \cite{WA82} (circles)
and the combined $D^\pm$ data from E769 \cite{E769}
(stars) are shown in Fig.\
3(a) and (b) respectively.
Because the data is given in arbitrary units, we
give a common normalization to
the calculations and data.  Figure\ 3(a) shows the
nonleading charm distributions for
$\pi^- p$ interactions at 340 GeV.  The
solid and dashed curves correspond
to fusion calculations using delta function
and Peterson function fragmentation respectively with the intrinsic charm
contribution of Eq.\ (8). The Peterson function result
lies well below the data, even
at moderate $x_f$. Figure\ 3(b) gives our
results for leading charm from $\pi^-
p$ interactions at the same energy,
including valence quark coalescence as well
as fragmentation of the intrinsic charm quarks as in Eq.\ (9).

Figure\ 3(c) compares our results with
the measured asymmetry at 250 (circles)
and 340 GeV (stars) as well as the
combined asymmetry from both experiments
(squares) \cite{WA82,E7692}.
 The steepest increase of ${\cal A}(x_f)$ arises from Peterson function
fragmentation (dashed curve) due to
its much softer nonleading distribution at
large $x_f$.  The solid curve shows
the delta function result.  Because of the
different nuclear target dependence
for the intrinsic charm contribution, the
leading charm effect diminishes in large nuclear targets.  The dot-dashed
curve shows delta function fragmentation without the average over nuclear
target to show that the relative target
dependencies of fusion and intrinsic
charm do not strongly affect ${\cal A}(x_f)$.
All the calculations based on
the intrinsic charm model reproduce the general
trend of the data.  Increasing
the coalescence contribution in Eq.\ (9)
by increasing $\xi$ would increase the
calculated asymmetry. The dotted curves in
Fig.\ 3(b) and (c) show the effect
of changing $\xi$ from $1/2$ to $9/10$ for a proton target.
The leading $D$ $x_f$ distribution is
marginally changed, but the asymmetry is
noticeably increased, bringing it into
better agreement with the high $x_f$ data.
This suggests that high statistics
single $D$ distributions combined with the
asymmetry data can fix relative
normalization of the coalescence contribution.
The leading $D^-$ rapidity
distribution from the two-component model is shown in Fig.\ 2(c) for both
$\xi = 1/2$ and $9/10$.
Note that the total charm rapidity distribution is
broader than the fusion result in Fig.\ 2(a), especially for $y>1$.

It should be emphasized that leading-twist QCD, including leading order
corrections, does not produce an asymmetry
between leading and nonleading charm
\cite{E7692,NDE}.  Both the intrinsic charm model and PYTHIA produce
an asymmetry with $x_f$.  In contrast to the PYTHIA predictions given in
\cite{WA82,E7692}, the intrinsic charm picture
produces a slightly negative
asymmetry for $x_f < 0.4$.
This is due to the difference between coalescence
and fragmentation at low $x_f$.
The $D$ meson $x_f$ distribution resulting
from fragmentation, calculated using Eq.\ (5),
peaks at $\langle x_f \rangle
\sim 0.25$, while the coalescence of the valence
quark with intrinsic charm
quarks, Eq.\ (6), produces leading charm with
$\langle x_f \rangle \sim 0.5$
\cite{VBH2}.  When $x_f > 0.4$, valence
quark coalescence dominates over all
other contributions, accounting for the rise in the calculated asymmetry.
Without this intrinsic charm coalescence
mechanism, {\it i.e.}\ $\xi = 0$,
there would be no asymmetry in this model.
The asymmetry produced by intrinsic
charm, shown in Fig.\ 3(c), corresponds to the single charm
production cross section
$\sigma_{\rm ic} \sim 1$ $\mu$b.  At higher energies, as the fusion cross
section increases, the relative intrinsic charm contribution to the charm
production cross section will diminish since $\sigma^{\rm in}_{\pi p}$ is
nearly constant with energy.
However, this will not affect the asymmetry at
large $x_f$ until the fusion cross section dominates the intrinsic charm
contribution.

The effects of intrinsic charm and coalescence occur dominantly at low
$p_T$ where the valence and charm quarks are aligned.  In order to
illustrate this effect, we note that the $p_T$ dependence of these models
is approximately \be \frac{dN_{\rm pf}}{dp_T^2} & \propto &
\frac{\alpha_s^2(p_T^2 + m_c^2)}{[p_T^2 + m_c^2]^2}
\, \, , \\ \frac{dN_{\rm
ic}}{dp_T^2} & \propto & \frac{\alpha_s^4(p_T^2
+ m_c^2)}{[p_T^2 + m_c^2]^4}
\, \, . \ee The higher--power falloff
in the $p_T$ dependence of intrinsic
charm reflects its higher--twist nature.  The higher power of $\alpha_s$
appears in Eq.\ (11) because the gluon
vertices with both the pion valence
quarks and the intrinsic $c \overline c$ pair must be included in the
calculation of the intrinsic charm
amplitude, leading to $\alpha_s^4$ in the
cross section.  In the parton fusion
model, the structure functions include the
coupling of gluons and quarks to the
incident hadrons, thus the fusion rate is
proportional to $\alpha_s^2$.

The E769 collaboration has also investigated
the $p_T^2$ dependence of the
asymmetry \cite{E7692}.  Figure\ 4 shows our
calculated $p_T^2$ dependence for
several regions of $x_f$.  Figure\ 4(c) compares
the combined $D^\pm$ $p_T^2$
distribution from E769 in the range
$0.1<x_f<0.7$ \cite{E769} with the model.
The solid and dashed curves indicate delta function and Peterson function
fragmentation respectively.
We have assumed that fragmentation only affects
the longitudinal momentum.
The calculations are weighted by the percentage of
the probability of intrinsic
charm production from fragmentation and coalesence
in each $x_f$ region.
Nonleading and leading charm calculations are shown in
Figs.\ 4(a) and (b).
The solid and dashed curves show our delta and Peterson
function results for $0<x_f<0.4$.
Peterson function fragmentation, with its
softer $x_f$ distribution, gives a somewhat larger $x_f$-integrated cross
section when $0<x_f<0.4$ than the delta function. Conversely,
when $0.4 < x_f < 1$, the
harder distribution of delta function
fragmentation (dot-dashed curves) gives
it a larger $x_f$-integrated cross section
than the Peterson function (dotted
curves), both for nonleading and leading charm.
The difference between the
delta and Peterson function predictions
at low $p_T^2$ in Fig.\ 4(b) is due to
the relative strength of valence quark
coalescence in the forward region.  As
$p_T^2$ increases, the difference caused
by the higher power falloff in $p_T^2$
is reduced primarily because of the higher
power of $\alpha_s$ in Eq.\ (11).
This difference will manifest itself most apparently in the asymmetry.

In Fig.\ 5(a), we compare the $x_f$-integrated
($0.1 < x_f < 0.7$) asymmetry of
E769 \cite{E7692} with our results.  There is no asymmetry for $x_f > 0$
because the integrated probabilities for leading and nonleading charm are
equal.  In the region covered by the data, there is a slight asymmetry,
shown by
the solid line for delta function fragmentation
and the dashed line for the
Peterson function.  Both curves are multiplied
by a factor of 50 to be visible.
The choice of the delta function leads to a negative asymmetry due to its
smaller integrated fragmentation component from Eq.\ (9).  The Peterson
function
tends to make the intrinsic charm fragmentation
component narrower, so that
most of its contribution is contained in the region $0.1 < x_f < 0.7$,
resulting in a positive ${\cal A}(p_T^2)$.
For $0 < x_f < 0.4$, the calculated
${\cal A}(p_T^2)$ is negligibly small and negative and is not shown.
When $0.4 < x_f < 1$, the
resulting ${\cal A}(p_T^2)$ is larger and
positive, as shown in Fig.\ 5(b).
The solid curve shows
delta function fragmentation, the dashed,
Peterson function.  Both decrease
strongly with $p_T^2$, due in part to the
different powers of the coupling
constant in the distributions.
Although the delta function is more physical in
low $p_T^2$ hadroproduction, the Peterson function should take over at
sufficiently large $p_T^2$ since
it describes jet fragmentation in $e^+e^-$
annihilation \cite{epem}.
A nonzero ${\cal A}(p_T^2)$ is more likely to be
found in the forward $x_f$ data at low $p_T^2$.

Recently, the Fermilab E791 collaboration has taken high precision charm
production data with a $\pi^-$
beam at 500 GeV on carbon and platinum targets
in the range $-0.1 < x_f < 0.8$ \cite{Carter}.  We have calculated the
$x_f$ and $p_T^2$ distributions
and the corresponding asymmetries for a proton
target for this experiment using GRV LO structure functions
and delta function fragmentation.  The fusion cross section increases
by a factor
of two between E769 and E791, decreasing the relative strength of the
intrinsic charm contribution by
$\sim 50$ \% in the higher energy experiment.
This reduces the asymmetry by
$\sim 10$ \% at $x_f \sim 0.8$ for E791.  The $p_t^2$ dependence of the
asymmetry is again the strongest
for the forward $x_f$ region, $0.4 < x_f <
0.8$.  (Below $x_f \sim 0.4$
${\cal A}(x_f)$ is again slightly negative; the
predicted asymmetry becomes positive for $x_f > 0.4$.)  We predict little
change in ${\cal A}(p_T^2)$ between E769 and E791.
The fusion cross section
will continue to increase with incident energy while ${\cal A}(x_f)$
will correspondingly decrease,
albeit more slowly than the intrinsic charm
contribution to the total cross section.\\[3ex]

\begin{center}
{\bf V.  Predictions for Leading Beauty Hadroproduction}\\[2ex]
\end{center}

We also use the two-component
model to predict $B$ meson distributions.  The
parton distribution functions
are more stable for $b \overline b$ production,
thus our calculations with the
GRV LO and DO 1.1 sets are very similar. We only
show delta function fragmentation since
there is less distinction  between the
Peterson and delta function predictions for $b$ quarks. One surprising
prediction of our model is that a larger
fraction of the inclusive $b$-quark
cross section is produced from intrinsic
beauty compared to the fraction of
open charm produced by intrinsic charm at
fixed energy. Of course, at fixed
values of $\tau = m_{Q \overline Q}/\sqrt{s}$, the fusion cross section
decreases by $ \sim (m_c/m_b)^2
(\alpha_s(m_{b \overline b})/\alpha_s (m_{c
\overline c}))^2$ between charm
and beauty production. The normalization of the
intrinsic heavy quark cross section contains factors of $\widehat{m}_Q^2
\alpha_s(m_{Q \overline Q})^4$
so that the probability for producing intrinsic
beauty relative to charm only
decreases as $(\widehat{m}_c/\widehat{m}_b)^2$
rather than $(\widehat{m}_c/\widehat{m}_b)^4$
as may be expected from Eq.\ (3).
 An additional decrease of
 $(\widehat{m}_c/\widehat{m}_b)^2$ is associated with
the resolving factor, $\mu^2/4\widehat{m}_b^2$.
We assume $\widehat{m}_b =
4.6$ GeV and take $\mu^2$ to be
independent of both the projectile and the
final state.  Thus, the intrinsic
beauty cross section is predicted to be \be
\sigma_{\rm ib} = P_{\rm ic} \left( \frac{\widehat{m}_c}{\widehat{m}_b}
\right)^2 \sigma_{\pi p}^{\rm in} \frac{\mu^2}{4 \widehat{m}_b^2} \left(
\frac{\alpha_s(m_{b \overline b})}
{\alpha_s(m_{c \overline c})} \right)^4 \, \,
, \ee on the order of 2-3 nb,
similar to the size of the fusion cross section
at the energies we investigate.
The nonleading and leading $B$ distributions
are similar to those given in Eqs.\ (8) and (9).
The fusion cross section is
expected to increase rapidly with energy but
$\sigma_{\rm ib}$ is proportional
to the inelastic cross section.
Therefore although $\sigma_{\rm ib}/\sigma_{b
\overline b}^{\rm total}$ is relatively
large for our calculated distributions,
it will decrease at higher energies.

No data on $B$ distributions or their
associated asymmetries have yet been
published.  We have calculated $B$ production in $\pi^- p$ interactions
at 250 and 340 GeV.  Our predictions of the leading and nonleading
distributions and the asymmetry between them is shown in Fig.\ 6.  The
solid lines in Figs.\ 6(a) and (c) illustrate the fusion mechanism
with delta function fragmentation.
The dashed curves show nonleading $B^+
(\overline b u)$
production with intrinsic $b$ quark fragmentation as in Eq.\ (8) and the
dot-dashed curves show leading $B^- (b \overline u)$'s as in Eq.\ (9).
The difference between the fusion and the nonleading $B$
distribution is larger than for charm since $\sigma_{\rm ic}$ and
$\sigma_{\rm ib}$ are coupled to the inelastic cross section.  As the
fusion cross section increases, $\sigma_{\rm ib}/\sigma_{b \overline
b}^{\rm total}$ is reduced, seen by comparing Figs.\ 6(a) and (c).
In Figs.\ 6(b)
and (d) we show the corresponding asymmetry calculations.  The
asymmetry is similar for charm and beauty because
the shape of the intrinsic distribution is only weakly dependent on the
heavy-quark mass.

Since fixed-target experiments can be
done at Fermilab with an 800 GeV proton
beam, we include $pp$ predictions for
$B$ production in Figs.\ 6(e) and (f)
using the same notation as above.
An intrinsic heavy quark state in the proton
has at least a five quark configuration, thus a leading $B^+$ or a $D^-$
produced by valence quark coalescence
has $\langle x_f \rangle \sim 0.4$ while
intrinsic heavy quark fragmentation gives
$\langle x_f \rangle \sim 0.2$, both
smaller than in a pion projectile.
The intrinsic heavy quark cross section with
proton projectiles is somewhat larger
since $\sigma_{pp}^{\rm in}$ is 40\%
larger than $\sigma_{\pi p}^{\rm in}$.
The $x_f$ distribution from fusion is
narrower in proton production, as shown
in Fig.\ 6(e).  A faster increase in
the asymmetry is expected (see Fig.\ 6(f)) since $\langle x_f \rangle$ is
reduced for the five-quark Fock state.

One interesting test of the extension of
this model to $B$ production is the
shape of the leading and nonleading distributions.
Since $\sigma_{\rm ib}$ is
comparable to the $b \overline b$ production cross section by fusion, a
parameterization of the $x_f$ distribution
as $(1-x_f)^n$ should give a harder
distribution and smaller $n$ than expected
from fusion production, especially
for $pp$ production, measurable at current energies.  As
the incident energy increases, $n$ should
increase as the fusion contribution
becomes dominant.  The difference between leading and nonleading $B$
distributions may become important for CP-violation studies where an
understanding of the symmetries of the $B$ hadroproduction cross section
is crucial.  This is particularly true for kaon beams in order to produce
$B_s$'s with a large Lorentz gamma factor.\\[3ex]
\eject
\begin{center}
{\bf VI. Conclusions}\\[2ex]
\end{center}

We have shown that the intrinsic charm model produces an asymmetry
between nonleading and leading charmed hadrons as a function of $x_f$ and
$p_T^2$.  The asymmetry, ${\cal A}$, is predicted to be largest at
low $p_T^2$ and to be an increasing function of $x_f$.
The model also accounts for the shapes of the charmed hadron
distributions in $x_f$.
It would be interesting to
look for an asymmetry between $\Lambda_c$ and $\overline \Lambda_c$
production both in $\pi^- p$ and $pp$ interactions.  Since both
$\Lambda_c$ and $\overline \Lambda_c$ contain a pion valence quark, the
asymmetry should be small.  However, the asymmetry could be large in $pp$
interactions where two of the proton valence quarks can coalesce with a
charm quark to produce leading $\Lambda_c$ while the $\overline
\Lambda_c$ should be centrally produced.
We also note that charmed-strange
mesons are produced at large $x_f$ by hyperon
beams \cite{WA62}, a leading
particle effect of the type studied here.

The asymmetry between $D^-$ and $D^+$ production
in $pp$ collisions should
also be checked.  If the two-component
model is correct, the asymmetry should begin to increase at lower $x_f$
than the calculated asymmetry with a pion projectile but should have
approximately the same shape.  We also predict that the $B$ and $D$
asymmetries should be similar.  An additional check on our model comes
from the shape of the $B$ distributions at large $x_f$, especially with a
proton beam where the intrinsic beauty contributions should
produce a broader distribution than
expected from leading twist fusion subprocesses.

\vspace{0.1in} \noindent {\bf Acknowledgements} We would like to thank J.
A. Appel, J. Hewett, and P. Hoyer for discussions.  We would also like to
thank T. Sj\"{o}strand for useful discussions
about the Lund model and PYTHIA
and for comments on the manuscript.

 \newpage

\begin{center}
{\bf Figure Captions}
\end{center}
\vspace{0.5in}

\noindent Figure 1.
The $x_f$ distributions for (a) $\pi^- p$ and  (b) $\pi^+
p$ interactions at 250 GeV and
(c) $\pi^- p$ interactions at 340 GeV calculated
from parton fusion.  The solid and dot-dashed curves show calculations
using GRV LO structure functions with delta function and Peterson
function fragmentation.
The dashed and dotted
curves illustrate the same results using DO 1.1 structure
functions.  The leading
$D^-$ data from the WA82 collaboration \cite{WA82} are
shown with the fusion calculation in (c).\\

\vspace{0.2in}

\noindent Figure 2.  The rapidity density of valence (solid curve) and
sea quarks (dotted) in the GRV
LO pion distributions \cite{GRVpi} are compared
with the rapidity density of the
produced charm quarks (dashed) in (a).  The
valence (solid) and charm (dashed) rapidity distributions in an pion
fluctuation into an intrinsic charm
state are shown in (b), along with the
resulting leading $D$ distribution from Eq.\ (6).  In (c) we show our
two-component model calculation including fusion with the leading charm
distribution from Eq.\ (9).  The model calculations with intrinsic charm
have been converted to rapidity.  All the calculations are for a 340 GeV
$\pi^-$ beam.\\

\vspace{0.2in}

\noindent Figure 3.  Our results for (a) nonleading charm and (b) leading
charm distributions in $\pi^- p$ interactions at 340 GeV and (c) the
asymmetry compared with the WA82 \cite{WA82} (circles) and
E769 \cite{E7692,E769} (stars) data.  The combined asymmetry from both
experiments is also shown (squares)
\cite{E7692}.  The calculations are with
GRV LO distributions using delta function (solid) and Peterson function
(dashed) fragmentation with the intrinsic
charm contributions to nonleading,
Eq.\ (8), and leading, Eq.\ (9), charm production.
The dotted curve in (b)
shows the leading $D$ distribution with $\xi
= 9/10$ in Eq.\ (9).  In (c), the
dashed curve is calculated with the Peterson
function and the solid curve with
delta function fragmentation.
Both are averaged over nuclear target.  The dot
dashed curve uses delta function
fragmentation and a proton target.  The dotted
curve shows the leading contribution
calculated with $\xi = 9/10$ in Eq.\ (9)
for a proton target.\\

\vspace{0.2in}

\noindent Figure 4.  Nonleading and
leading charm $p_T^2$ distributions are
shown in (a) and (b) for 250 GeV $\pi^- p$
interactions.  The solid and dashed
curves are calculated using delta function
and Peterson function fragmentation
in the $0 < x_f < 0.4$ interval while the
dot-dashed and dotted curves are
calculated for $0.4 < x_f < 1$.
The combined $D^\pm$ $p_T^2$ data from E769
\cite{E769} is compared to our
calculation integrated over $0.1 < x_f < 0.7$ in
(c).  The solid and dashed curves use delta function
and Peterson function fragmentation respectively.\\

\vspace{0.2in}

\noindent Figure 5.  The $x_f$-integrated
($0.1 < x_f< 0.7$) asymmetry of E769
\cite{E7692} is shown in (a).  Our results,
using delta (solid) and Peterson
(dashed) fragmentation functions, are multiplied by a factor of 50.  The
asymmetry in the forward region,
$0.4 < x_f < 1$, is shown in (b). The solid
and dashed curves are again our results for the delta and Peterson
fragmentation functions.\\

\vspace{0.2in}

\noindent Figure 6.  Our predictions of leading and nonleading $B$
distributions are shown in (a), (c), and (e).  The solid lines illustrate
the fusion mechanism.  The dashed curves show nonleading $B$ production
with intrinsic $b$ quark fragmentation and the dot-dashed curves are
leading $B$'s ($B^- (b \overline u)$
for $\pi^- (\overline u d)$ and $B^+ (u
\overline b)$ for $p (uud)$).  In (b), (d), and
(f), we show the corresponding asymmetry calculations.

\end{document}